\documentclass[twocolumn,showpacs,preprintnumbers,amsmath,amssymb]{revtex4}
\usepackage[utf8]{inputenc}
\usepackage{graphicx}% Include figure files
\usepackage{dcolumn}% Align table columns on decimal point
\usepackage{bm}% bold math

\begin{document}

\title{Calculation of density of states of transition metals: from bulk sample to nanocluster}% Force line breaks with \\

\author{Petr V.  Borisyuk}
\affiliation{%
National Research Nuclear University MEPhI Kashirskoye shosse 31, Moscow, 115409, Russian Federation 
}%

\author{Oleg S. Vasiliev}
\affiliation{%
National Research Nuclear University MEPhI Kashirskoye shosse 31, Moscow, 115409, Russian Federation 
}%

\author{Yaroslav V. Zhumagulov}
\affiliation{%
National Research Nuclear University MEPhI Kashirskoye shosse 31, Moscow, 115409, Russian Federation 
}%

\author{Vladimir A. Kashurnikov}
\affiliation{%
National Research Nuclear University MEPhI Kashirskoye shosse 31, Moscow, 115409, Russian Federation 
}%

\author{Andrey V. Krasavin}
\affiliation{%
National Research Nuclear University MEPhI Kashirskoye shosse 31, Moscow, 115409, Russian Federation 
}%

\author{Uliana N. Kurelchuk}
\affiliation{%
National Research Nuclear University MEPhI Kashirskoye shosse 31, Moscow, 115409, Russian Federation 
}%

\author{Yuriy Yu. Lebedinskii}
\affiliation{%
National Research Nuclear University MEPhI Kashirskoye shosse 31, Moscow, 115409, Russian Federation 
}%
\date{\today}% It is always \today, today,
             
\begin{abstract}
The technique is presented of restoring the electronic density of states of the valence band from data of X-ray photoelectron spectroscopy. The originality of the technique consists in using a stochastic procedure to solve an integral equation relating the density of states and the experimental X-ray photoelectron spectra. The results are presented for bulk sample of gold and nanoclusters of tantalum; the possibility of using the results to determine the density of states of low-dimensional structures, including ensembles of metal nanoclusters, is demonstrated.
\end{abstract}
\maketitle

\section{Introduction}
\label{Introduction}
Low-dimensional structures such as thin films, nanotubes, metal nanoclusters, which are situated in an intermediate position between single atoms and bulk materials, possess unique physical properties that can separate them in a special area of condensed matter physics. Studying of the properties of such structures attracts now an increased interest both from a fundamental point of view, and in connection with numerous applications. In many ways, the practical use of low-dimensional structures, and those consisting of transition metals in particular, is based on the features of their electronic properties due to strong electron correlations \cite{1}. The necessity of the accounting of electron correlations leads to considerable complexity of the theoretical description of transition metals, even for macroscopic systems. In the case of transition metal structures of reduced dimensionality, the description of the electronic properties is an even more difficult task. For this reason, the development of experimental and numerical methods for the study of the electronic properties of nanoscale structures of transition metals is an urgent task.

X-ray photoelectron spectroscopy (XPS) \cite{2} allows studying the evolution of the electronic properties of low-dimensional structures depending on their size. This is because the thickness of a layer analyzed by XPS and determined by an average path length of a photoelectron in the sample material, is 5-30 $\AA$. This fact makes the XPS technique sensitive to the surface density of electronic states, and gives the opportunity to study physical and chemical properties of low-dimensional samples. The advantage of this technique is also the possibility of direct measurement of the density of electronic states (DOS) of the valence band. The smaller is the instrumentation broadening, which is determined mainly by nonmonochromaticity of the X-ray source and by the operation mode of the energy analyzer, the more precise is the determination of the features of DOS. Modern X-ray photoelectron spectrometers equipped with a monochromator have the value of the broadening ranged from 0.4 to 0.6 $eV$. This broadening leads to smearing of DOS and usually prevents direct study of the spectra. On the other hand, alternative methods for studying the electronic properties, such as the method of scanning tunnel spectroscopy, the method of optical absorption spectroscopy, positron annihilation technique, techniques related to the de Haas-van Alphen effect, the method of ultraviolet electron spectroscopy (UPS), and the method of electron energy loss spectroscopy (EELS), are much more complex in terms of the interpretation of the spectra. For instance, the UPS technique, which is usually used to obtain spectra of the valence band, reduces the instrumentation broadening up to several $MeV$, but the signal intensity in this case is proportional to the convolution of the density of occupied and free electron states. A similar problem exists in other mentioned methods, and in the case of EELS one should also take into account the plasmon and phonon contributions to the spectra.  

In this paper, we propose the technique of restoring the electronic density of states of X-ray photoelectron spectra using a numerical stochastic procedure, which accurately takes into account the instrumentation broadening. XPS spectra of a sample obtained in the regions of valence states and core levels allow the obtaining the instrument function of the spectrometer. For this purpose the core level line is used; the parameters of its broadening are derived from the approximation of the experimentally measured spectrum. Further, the stochastic procedure is applied to restore DOS of nanoclusters by solving an integral equation. This technique was used to calculate DOS of bulk gold and nanoclusters of tantalum. To confirm the accuracy, the obtained DOS for bulk gold was compared with DOS calculated from density functional theory (DFT).

\section{XPS technique}
\subsection{Shape of XPS lines}
The shape of XPS lines in the scale of the binding energy $E$ with respect to the Fermi energy is defined by the following convolution \cite{2}:
\begin{equation}
\label{eq:1}
I(E)=I_1(E)*I_2(E)*I_3(E)*I_4(E),
\end{equation}
where
\begin{equation}
\label{eq:2}
I_1(E)\sim \delta (E-E_0)
\end{equation}
is the spectrum of the core level with the binding energy $E_0$ without broadening factors (the lifetime of a hole on the core level is infinite; the instrumentation broadening is equal to zero);
\begin{equation}
\label{eq:3}
I_2(E)\sim \frac{\gamma }{\sqrt{E^2+{\gamma }^2}}
\end{equation}
is the broadening of the spectral line due to the finite lifetime of a hole on the core level $\tau $, the natural broadening $\gamma \sim \hslash /\tau $;
\begin{equation}
\label{eq:4}
I_3(E)\sim 
\begin{cases}
E^{\alpha -1},& \text{if } E>0\\
0,                  & \text{if } E<0
\end{cases}
\end{equation}
is the broadening of the spectral line associated with the many-electron effects; $0\le \alpha <1$ is Anderson parameter of singularity \cite{3};
\begin{equation}
\label{eq:5}
I_4(E)\sim {\mathrm{exp} (-\frac{E^2}{2{\sigma }^2})\ }
\end{equation}
is Gaussian intensity distribution determined by the instrumental broadening, ${\sigma }^2=\frac{W^2_{hv}+W^2_{sp}}{4{\mathrm{ln} 4\ }}$, where $W^{\ }_{hv}$ is line width of X-ray radiation; $W^{\ }_{sp}$ is the instrumentation broadening.
For valence electrons the lifetime of a hole tends to zero, therefore, the contribution of the natural broadening can be neglected in \eqref{eq:1}. Then, the XPS spectrum of the valence band can be written as 
\begin{equation}
\label{eq:6}
I_{val}(E)=A_{val}(E)*I_3(E)*I_4(E),
\end{equation}
where $A_{val}(E)$ is the level density of valence states. Thus, with known functions $I_3(E)$ and $I_4(E)$ one can obtain the behavior of DOS near the Fermi level.

\subsection{Restoration of DOS from XPS spectra}
The convolution of functions $I_3(E)$ and $I_4(E)$ in \eqref{eq:6} is a instrument function $H(E)$ defined by technical characteristics of the spectrometer and many-electron processes:
\begin{equation}
\label{eq:7}
H(E)=I_3(E)*I_4(E).
\end{equation}
In this case, experimental XPS spectra of the core line $I_{core}(E)$ and the valence band $I_{val}(E)$ can be written as a system of Fredholm integral equations of the first kind: 
\begin{equation}
\label{eq:8}
\begin{cases}
I_{core}(E)=\int^{+\infty }_{-\infty }{H(E-E^{'}) I_2(E^{'})       dE^{'}};\\
I_{val}(E)  =\int^{+\infty }_{-\infty }{H(E-E^{'}) A_{val}(E^{'}) dE^{'}}.
\end{cases}
\end{equation}
It follows from \eqref{eq:8} that after restoring the instrument function $H(E)$ from XPS spectrum $I_{core}(E)$ of the core line with known collision breadth, one may restore DOS in the valence band $A_{val}(E)$. In this case the problem of calculation the parameters $\alpha $, $\sigma $ and $\gamma $ can be reduced to the approximation of the core XPS line by the least square method.

The task of restoring the electronic density of states $A_{val}(E)$ is the ill-posed problem of solving the system of integral equations \eqref{eq:8} with the obtained instrument function. Among many approaches of numerical solution of such ill-posed problems, one of the most effective is the "algorithm of random rectangles" first proposed in \cite{4}. The basic idea of the algorithm is to approximate $A_{val}$ with a piecewise constant function 
\begin{equation}
\label{eq:9}
{\tilde{A}}_{val}(E)=\sum{{\chi }_{c,w,h}(E)},
\end{equation}
where
\begin{equation}
\label{eq:10}
\chi _{c,w,h}(E)=
\begin{cases}
h,& \text{if } c-\frac{w}{2} < E < c+\frac{w}{2}; \\ 
0,&   \text{otherwise }; 
\end{cases}
\end{equation}

parameters $c$, $w$, $h$ are used for the center, the width, and the height of the rectangle, respectively. The algorithm is based on minimizing the deviation between the input signal $I(E)$ (experimental XPS spectrum) and approximated $\tilde{I}(E)=\int^{\infty }_{-\infty }{H(E-E^{'})}\tilde{A}(E^{'})dE^{'}$. The process of generating the solution is a random selection of the initial configuration and its subsequent optimization to minimize deviations from the exact solution by changing the values of parameters and the number of rectangles. Technically, any solution of the integral equation can be parameterized with any desired accuracy with the use of such procedure. Various functionals can be used to evaluate the deviation; in this work the following functional was used \cite{5}:
\begin{equation}
\mathrm{\Delta }=\int^{+\infty }_{-\infty }{{(I(E)-\tilde{I}(E))}^2dE}.
\end{equation} 
The test results are shown for bulk gold as a reference material for XPS measurements. 

\subsection{Experiment details}

The analysis of XPS spectra of bulk gold was carried out in the UHV chamber ($1\times {10}^{-10}$ Torr) of the Surface Analysis System Theta Probe (Thermo Scientific). The sample was cleared before measurements by etching with ${\mathrm{Ar}}^+$ ions with the energy of 2 KeV during 2 min. Photoelectron spectra of the core and valence levels were obtained using the monochromatic AlK$\alpha_{1,2}$ ray source (1486.6 eV). The kinetic energy of electrons was measured in increments of 0.05 eV, which ensured the measurement error at the level of $\pm$0.03 eV. The calibration of the binding energy scale was carried out by the position of Au$4f_{7/2}$ spectral line of bulk gold. The binding energy of Au$4f_{7/2}$ level of chemically pure gold is 83.96$\pm$0.01 eV \cite{6}. The spectra were recorded with the pass energy of the spectrometer of 20 eV. The validity of the experimental data is confirmed by reproducibility of the spectra for series of samples measured under identical conditions.

\section{ Results and discussion}
Typical X-ray photoelectron spectra of the core levels of bulk gold are presented in Fig. 1. The observed Au$4f_{7/2}$ doublet is due to the spin-orbit splitting. The position of Au$4f_{7/2}$ line corresponds to the binding energy of 87.67 eV and to the difference of 3.71 eV between the lines of the doublet. It is consistent within the measurement error with the known value of the energy of the spin-orbit splitting of the $4f$ level for bulk gold \cite{6}.  The instrument function $H\left(E\right)$ restored by fitting the experimental Au$4f_{7/2;5/2}$ spectrum, is shown in Fig. 2.

The following values defining the shape of the XPS spectrum were obtained from the approximation: $\gamma =0.23\ eV;\ \alpha =0.04;\ \sigma =0.23\ eV$. These values are in agreement with known data.

\begin{figure}
\includegraphics[width=8cm]{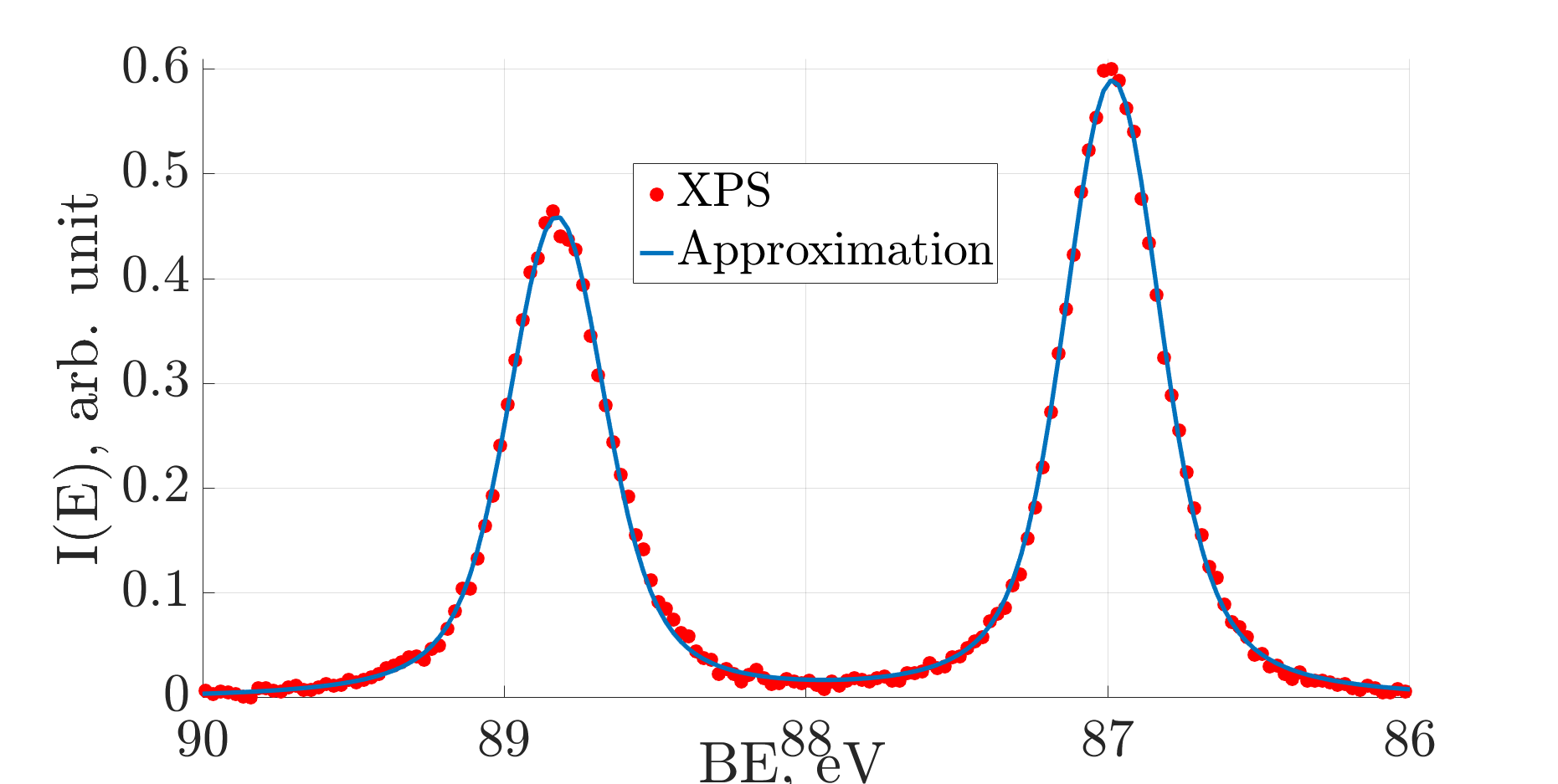}
\caption{\label{fig:1} Experimental XPS spectrum of Au$4f_{7/2;5/2}$ doublet and its approximation \eqref{eq:1}}
\end{figure}

\begin{figure}
\includegraphics[width=8cm]{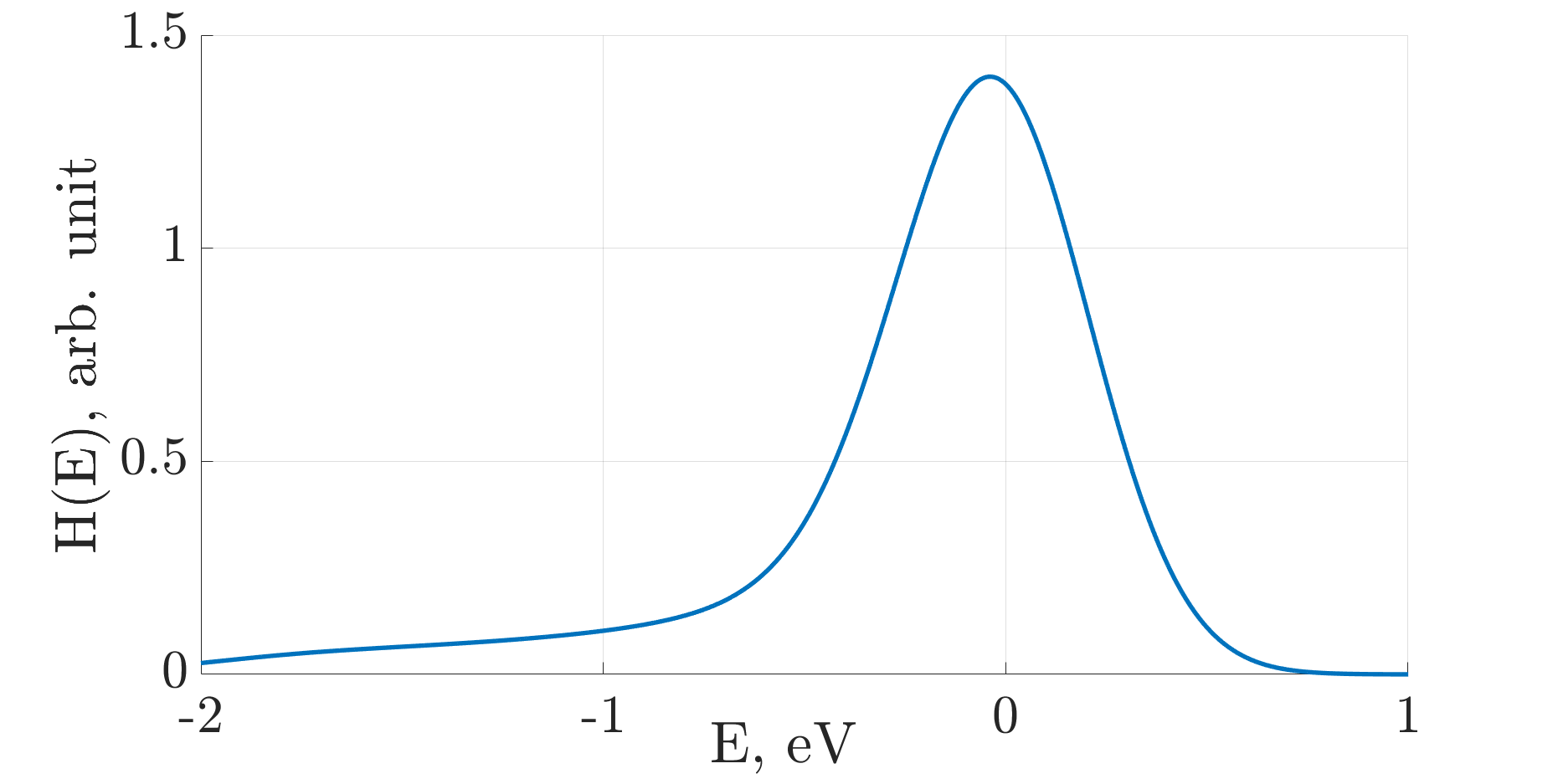}
\caption{\label{fig:2}  The restored instrument function $H\left(E\right)$}
\end{figure}

\begin{figure}
\includegraphics[width=8cm]{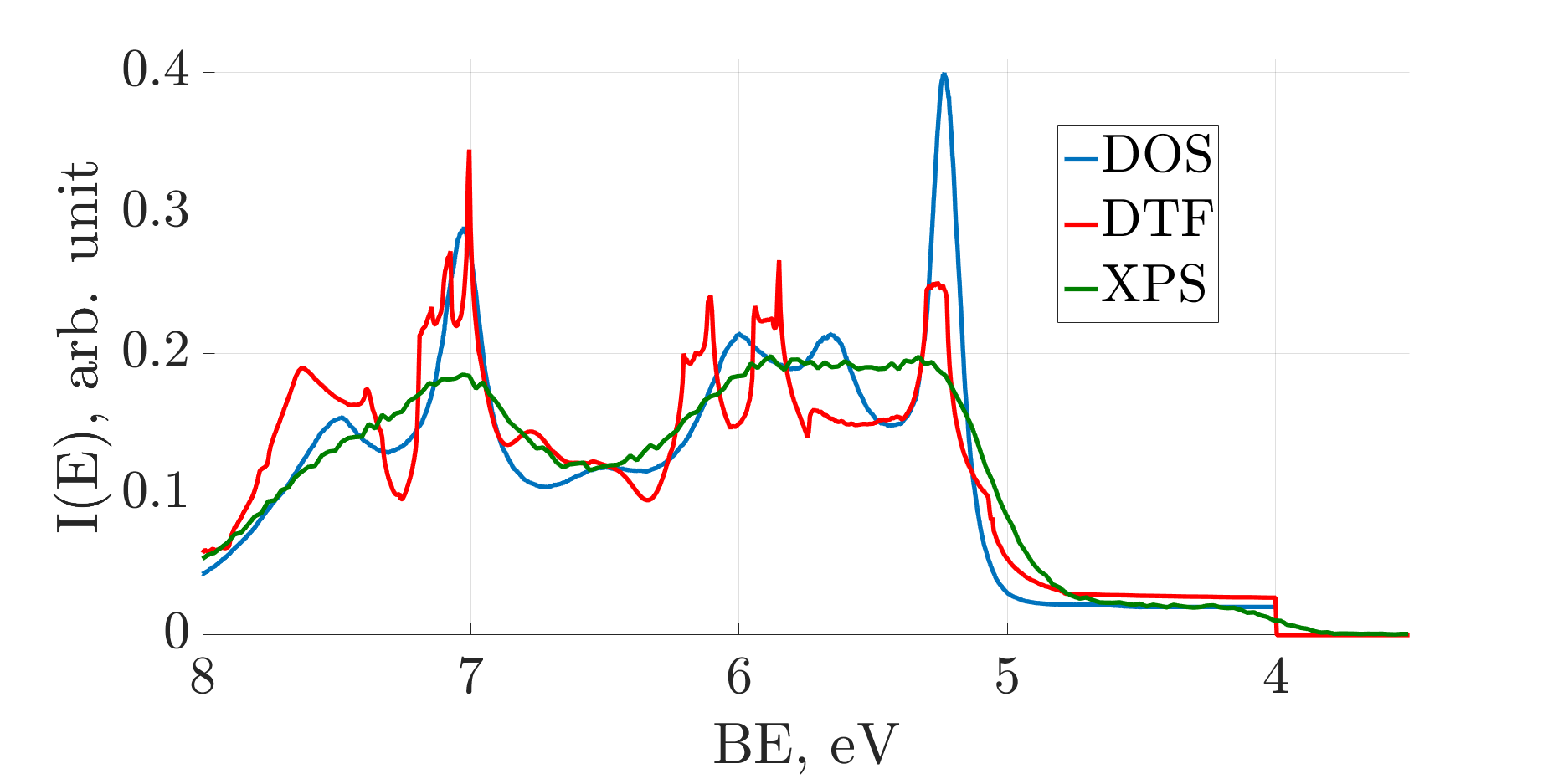}
\caption{\label{fig:3}  The experimental XPS spectrum of the valence band, the restored DOS, and DOS calculated by DFT}
\end{figure}

The obtained instrumental function $H(E)$ was used for the calculation of DOS for bulk gold in accordance with the system of integral equations \eqref{eq:8}. Fig. 3 shows the calculated DOS in comparison with DOS obtained by DFT method. DFT calculations were conducted in the generalized gradient approximation with the Perdew--Burke--Ernzerhof exchange-correlation potential taking into account the spin-orbit interaction and non-collinear magnetism \cite{7}, with zero initial magnetization; Quantum Espresso 5.4.0 software package \cite{8} was used. To take into account the spin-orbit interaction fully relativistic ultrasoft pseudopotentials have been used, generated in Atom package (Quantum espresso) \cite{9}. It can be seen that the features of the restored DOS correlate with DFT calculation, in spite of the rather complicated form of the spectrum. 

As is known, the valence band of gold is determined by the outer $5d^{10} 6s^{1}$ electrons. There is a step on the spectrum (see Fig. 3) in the vicinity of zero binding energy values (the Fermi level). This step corresponds to the half-filled $6s$ shell. The blurring of this step determines the instrumentation broadening, which proved to be equal to 0.54 eV; this agrees within the error with the certified value of the spectrometer for the selected mode. By increasing the binding energy, sharp peaks are observed, associated with the contribution of \textit{d} electrons. The peaks in the energy range of 2 -- 4 eV, according to DFT calculations, correspond to the hyperfine structure of $5d_{5/2}$ electrons, and the peaks in the energy range of 5-8 eV correspond to $5d_{3/2}$ electrons. The presence of distinct features in DFT calculations at energies of 2.5 eV and 6.0 eV is in agreement with the shape of DOS recovered from the experimental spectra. In the XPS spectrum, the identification of these features is difficult because of the instrumental broadening. The correlation between the calculated DFT-data and the restored DOS allows us to affirm the adequacy of the models used. 

\section{Conclusions}
In this paper, we proposed the method of obtaining information about the electronic structure of the occupied states near the Fermi energy. The method uses experimental XPS spectra and stochastic numerical procedure to restore the density of states. The key feature of the method, distinguishing it from first-principles calculations, is that it is not necessary to take into account the crystal structure of nanoclusters, only the XPS spectra of the core levels are needed to restore the instrument function. This advantage allows the use of the method to determine the dependence of the properties of low-dimensional structures on their characteristic size.

\section*{Acknowledgements}
The work was supported by Russian Science Foundation (project \# 16-19-00168).

\end{document}